\documentstyle[epsfig,aps,10pt,prb]{revtex}

\def\hv{{\bf h}}

\begin{document}
\title{Teaching statistics in the physics curriculum: \\
Unifying and clarifying role of subjective probability}

\author{Giulio D'Agostini \\}
\address{\it Dipartimento di Fisica dell'Universit\`a ``La
Sapienza''
\\ and Istituto Nazionale di Fisica Nucleare (INFN),
P. le Aldo Moro 2, I-00185 Roma, Italy}

\maketitle
\centerline{E-mail: dagostini@roma1.infn.it.}

\begin{abstract}
\vspace{-6.1cm}
\begin{flushleft}
{\large Roma1 N.1108} \\
\vspace{-2.5mm}
\mbox{} \\
{\large  Luglio 1999} \\
\vspace{-2.5mm}
\mbox{} \\
{\large \tt physics/9908014}
\end{flushleft}
\vspace{4.2cm}

Subjective probability is based on the intuitive idea that
probability quantifies the degree of belief that an event will
occur. A probability theory based on this idea represents the
most general framework for handling uncertainty. A brief
introduction to subjective probability and Bayesian inference is
given, with comments on typical misconceptions which tend to
discredit it and comparisons to other approaches.

\end{abstract}

\section{Introduction}
\footnotetext{~To appear in the special theme issue 
of the {\it American Journal of Physics} on {\sl 
Thermodynamics, Statistical Mechanics, and Statistical Physics}, 
H. Gould and J. Tobochnik  eds., December 1999 (see 
http://stp.clarku.edu/\\ajp\_contributors.html).}
Physics students encounter concepts of probability and statistics
several times during their studies, usually in laboratory classes
when the treatment of measurement errors is introduced and in
statistical physics and quantum mechanics courses. Some universities
also provide specialized courses on probability and statistics.
However, there is a general consensus that the standard
understanding of statistics is insufficient and
confused.\cite{maxent98} In my opinion, the main reason for this
unsatisfactory situation is that the fundamental issue concerning
the concept of probability, which should precede any exposition of
probability, is not treated with due care. 

The purpose of this article is to introduce probabilistic 
reasoning from the point of view of subjective probability, on
which Bayesian statistics is based. The choice of name is due to
the key role played by Bayes' theorem in updating probability in
the light of new information.

\section{Subjective probability}

We often find ourselves in a status of uncertainty about {\it
events} which might occur. For example, a tossed coin would result
in heads or tails (two possible events). Or, given
$N$ molecules at equilibrium in a box, we might be interested in
the number of molecules at a given instant which are present in a
sub-volume of the box ($N+1$ events).

In general, we know that all events do not have the same chance of
occurring. Consider two events $E_1$ and $E_2$. Stating that $E_1$
is more probable than
$E_2$ ($P(E_1) > P(E_2)$) means that {\it we} consider 
$E_1$ to be more likely to occur than $E_2$. This statement is
no more than the concept of probability that the 
human mind has developed naturally to
classify the plausibility of events under conditions of 
uncertainty.\cite{probabile} In other words, probability is 
related to the ``degree of belief in the occurrence of an 
event.''\cite{deFinetti} The 
usual definition of subjective probability one finds in 
introductory books is {\it ``the
degree of belief that an event will occur.}''\cite{will}

This definition of the concept of probability is not bound to a
single evaluation rule, and there are many ways to obtain $P(E)$.
The assessment could be based on symmetry considerations, past
frequencies, Monte Carlo simulations, complicated theoretical
formulae, or Bayesian inference. What matters is that the meaning
is the same in all applications, and is independent of the
method of evaluation. For example, if we state that the probability of a 
$Z^\circ$ boson decaying to an $e^+e^-$ pair is 3.3\%, and that of
observing 5 heads after 5 fair coin tosses is 3.1\%, it means that
we are slighter more confident that a $Z^\circ$ will decay into
$e^+e^-$ than five tossed coins give all heads.

We also note that probability assessments depend on who (the
``subject'') does the evaluation and, more precisely, on the status
of the information that the subject holds at the moment of the
assessment. Therefore what matters is always conditional
probability, conditioned by the status of information $I$, that is,
$P(E\,|\,I)$ is to be read ``the probability of $E$ {\it given}
$I$.'' As a consequence, several persons might have simultaneously
different degrees of belief on the same event, as is well known to
poker players.

Subjective probability tends to disturb scientists, who pursue the
ideal of objectivity. But, rigorously speaking, an objective
knowledge of the physical world is impossible, if ``objective''
stands for something which has the same logical strength as a
mathematical theorem.\cite{Hume} Nevertheless, if rational people
share the same information, the ideal of objectivity is recovered 
through intersubjectivity. 

Subjective probability does not imply that we may believe whatever
we like, for example, flying horses or speaking dogs. I can
{\it imagine} a flying horse as a combination 
of concepts that I have
from my experience, but nevertheless, 
I do not {\it believe} flying
horses to exist.\cite{Hume1} 
There is a crucial ingredient of the
subjective approach which forces people to make probability
assessments that correspond effectively to their beliefs. This
ingredient is the so-called {\it coherent bet}.\cite{deFinetti} If
we consider an event to be 50\% probable, then we should be
ready to place an even bet on the occurrence of the event or on its
opposite. However, if someone is ready to place the bet in one
direction but not in the other direction, it means that this person
thinks that the preferred direction is more probable than the other,
and then the 50\% probability assessment is {\it incoherent}, that
is, this person is making a statement which does not correspond to
his belief.

Even if an event and its opposite ($\overline{E}$) are not
equiprobable, a bet can still be arranged if the odds are fixed
proportionally to the beliefs on the two events:
$\mbox{odds ratio}(E:\overline{E}) = P(E):P(\overline{E})$.
Therefore, if someone considers a 2:1 bet in favor of
$E$ to be fair, it means that that person judges
$P(E)=2/3$. Coherence prevents people from arbitrary probability
assessments.\cite{torta}

A coherent bet has to be considered {\it virtual}. For example, a
person might judge an event to be 99.9999\% probable, but
nevertheless refuse to bet 
\$999999 against \$1, if \$999999 is the order of magnitude of
the person's resources. Nevertheless, the person might be convinced
that this bet would be fair if he had an infinite budget. This
remark teaches us that probability assessments should be kept
separate from decision issues. The latter can be more complicated,
because decisions depend not only on the probability of the event,
but also on the subjective importance of a given amount of money.

The first consequence of coherence is that probability assessments
can be exchanged among rational people, with the guarantee that
everybody is talking about the same thing, although the evaluations
might differ due to a different status of information. The second
important consequence\cite{deFinetti} is that it is possible to
derive from the requirement of coherence the basic rules or axioms
of probability.\cite{axioms} We will not give the derivation here,
but simply summarize the well known rules: 
\begin{eqnarray}
&& 0\le P(E) \le 1 \label{eq:basic1}\\
&& P(\Omega) = 1 \label{eq:basic2}\\
&&
P(E_1\cup E_2) = P(E_1)+P(E_2) \quad
\mbox{if} \quad E_1\cap E_2 = \emptyset, \label{eq:basic3}
\end{eqnarray}
where $\Omega$ and $\emptyset$ stand for the certain and the
impossible event, respectively, $\cap$ represents the 
{\it logical product} (also known as ``AND''), and 
$\cup$ the {\it logical sum} (``OR'').
 
Another important relation which can be derived from coherence
is the relation between joint probability and conditional
probability:
\begin{equation}
P(A\cap B) = P(A\,|\,B) \, P(B) = P(B\,|\,A)\, P(A)\,,
\label{eq:joint}
\end{equation}
where $P(A\,|\,B)$ is the probability of the event $A$ under
the hypothesis that $B$ is true. 
In the axiomatic approach
Eq.~(\ref{eq:joint}) arises from the ``definition'' of conditional
probability,\cite{conditional}
that is,
\begin{equation}
P(A\,|\,B) = \frac{P(A\cap B)}{P(B)}\,. \hspace{1.0cm} (P(B)\ne 0)
\label{eq:cond} 
\end{equation} 

Because the basic rules of probability, Eqs.~(1)--(4), derived from
coherence are the same as those introduced in the axiomatic
approach, all other probability rules, as well as the probability
calculus, are the same. But the subjective approach does more.  It
guarantees that if the numbers we use at the beginning of a
calculation are coherent degrees of beliefs, the result also has to
be interpreted as a degree of belief, necessarily following from
the initial ones.  For example, if we believe that a coin has a
60\% chance to give heads, then we implicitly attribute a 23\%
chance to 5 independent tosses of that coin to produce exactly 3
tails.\cite{coins}

\section{Interplay of subjective probability with combinatorial
and frequency based evaluations}\label{sec:interplay}

It is not difficult to realize that the usual definitions of
probability in terms of the ratio of favorable to possible cases,
or of successes to trials,
cannot define the concept of probability, because they are based on
the primitive concept of equiprobability (see for example
Ref.~\ref{bib:dagocern}). Nevertheless, in the subjective approach
these ``definitions'' can be easily recovered as useful evaluation
rules.\cite{Hume2} 

The use of combinatorial evaluation is rather obvious, and the
common urn and dice  problems yield ``objective'' answers, in the
sense that all reasonable people will agree. Given $N_W+N_B$
indistinguishable  white and black balls in an urn, there is no
reason to consider a particular ball to be more likely to be
extracted (otherwise, we should bet more money on that ball than
on the others). Then, as a straightforward application of
Eqs.~(2)--(3), we find
$P(\mbox{white}) = N_W/(N_W+N_B)$ and 
$P(\mbox{black}) = N_B/(N_W+N_B)$. Sometimes urn problems are
considered to provide a reference (or calibration) probability. If
I assign 80\% probability to the event $E$, it means that I am as
confident that this event will result as I am confident of
extracting a white ball from an urn which contains 100 balls, 80
of which are white. Everybody understands how much I am confident in
$E$, independently of what $E$ might be.
 
More generally, combinatorics (for countable events)  and measure
theory (when events form a continuum class) are just mathematical
tools of probability theory, if the elements of the relevant space
are judged to be  equiprobable. This point of view is the exact
opposite and, in my opinion, more physical than that stated in many
books on mathematical or statistical physics  (for example,
``probability theory \ldots is certainly a branch of analysis and
in a narrow sense a branch of measure theory. Its most rudimentary
parts are rooted in combinatorics.''\cite{Kac})

The frequency based definition of probability needs
a more extensive discussion. 
Empirical frequencies can be used to evaluate probability by
stating that we believe that what has happened more often in the
past will happen more probably in the future.\cite{Hume} This 
simple evaluation rule is applicable if there are no other
relevant pieces of information to take into account. Past
frequencies can also be used in a more formal way, together with
other information, by applying
Bayesian inference, which will be 
introduced below. 
In general, the value of a
probability will not be exactly equal to the relative frequency.
Only when the number of past experiments is very large will the
results of Bayesian and empirical frequency evaluations converge to
the same value. An example will be given in
Section~\ref{sec:inference} which shows quantitative disagreement
between the two methods for a finite number of measurements. 

Let us see more carefully how
frequentists make use of their probability definition. It is clear that
the use of past frequencies to evaluate probability relies on a belief
that the measurements were done under the same conditions (of
equiprobability) and that the relative frequency has approached a limit.
Thus, it is not correct to say that the frequentist
approach is free of subjective ingredients. Moreover, can frequentists
assess that, for example, the probability of extracting a white ball from
an urn which contains 70 white balls and 30 black balls is 70\%?
Apparently they cannot, unless they have done an experiment to
``measure'' the probability from a long series of experiments.
Nevertheless, they do, using the following type of reasoning.\cite{Polya}

\begin{enumerate}

\item We first say that ``we see no reason why one ball should be
preferred to another.''\cite{Polya} (The expression ``equally
probable'' is avoided, but the meaning is exactly the same.)

\item ``We naturally expect that, in the long run, each ball will be
drawn approximately equally often.''\cite{Polya} It follows that
the frequency of each ball is expected to be approximately similar
and the frequency of white balls is proportional to their number in
the box.

\item Finally, we ``expect'' a relative frequency approximately
equal to the proportion of white balls in the box. Therefore,
the probability is equal to the proportion of white balls. 

\end{enumerate}

In some texts (see for example, Ref.~\ref{bib:Reif}), ``a priori
probabilities'' are introduced by an {\it ad hoc} postulate;
``\ldots once the basic postulate has been adopted, the theory of
probability allows the theoretical calculation of the probability
of the outcome for an experiment.''\cite{Reif} But it is clear
that in this context ``postulate'' is nothing but ``belief,'' but
it sounds nobler.
 
In the subjective approach the terms of the problem are better
defined and have a closer correspondence to intuitive concepts. In
particular, a clear distinction is made between the following three
ingredients which enter statistical considerations: past frequency,
probability, and future frequency (``future'' refers to unknown
results, not necessarily occurring later in time.\cite{will}) We
now analyze the same example from the subjectivist perspective.

\begin{enumerate}

\item Given our status of knowledge, we have no reason to believe
that one ball will be extracted more likely than the others
(otherwise, we should be ready to bet more money on that
particular ball). Therefore, we judge them all equally probable and,
applying the basic rules of probability, we assign 70\% probability
to white. The 70\% probability has a precise and intuitive meaning
by itself, as a degree of belief of the result of any
extraction. There is no need to think about a statistical ensemble
of many such experiments. This reasoning might sound similar to the
first point of the frequentist's perspective. But in
the frequentist approach the reasoning is very convoluted, 
because they do not speak about the probability of individual 
events, but only of ``random mass phenomena,''\cite{vonMises}
as illustrated in Ref~\ref{bib:Polya}.

\item Nevertheless, we can always think of $N$ experiments with 
the same urn, reintroducing the ball after each extraction, or,
more generally, of $N$ independent events, each of which is
believed to occur with 70\% probability. The relative frequency of
the white balls, $f_W$, is an uncertain number with $N+1$
possibilities, to each of which we attribute a degree of belief,
$P(f_W)$, a consequence of the degree of belief of the individual
event ($p=70\%$) and of the believed independence of the $N$ events:
\begin{equation}
P(f_W) = 
 \left(\!\!\begin{array}{c} N \\ 
 N f_w
 \end{array}\!\!\right)
\,p^{N f_w }\,(1-p)^{N(1-f_w)}\,.
\label{eq:pfreq}
\end{equation}
The expected value and standard deviation of the frequency
are $E(f_W)= p$ and
$\sigma(f_W) = \sqrt{p\,(1-p)}/\sqrt{N}$. 
These two quantities are related to the concepts of (probabilistic)
{\it prevision}\cite{prevision} and of (standard) {\it
uncertainty} of the prevision, respectively. When we consider a
very large
$N$, we judge that it is very unlikely to obtain a value of the
relative frequency that differs more than 70\%, as is born out by
Eq.~(\ref{eq:pfreq}). This result is precisely what is expected from
the law of large numbers, expressed by Bernoulli's
theorem,\cite{more} a consequence of Eq.~(\ref{eq:pfreq}).

\end{enumerate}

Let us summarize the subjectivist
point of view about past frequency, probability, and future
frequency.\cite{teaching}
Past frequency is experimental data, something that
happened with certainty and to which the category of probability no
longer applies. Probability is {\it how much we believe} that
something will happen, taking into account all available
information about the event of interest, including, if they are
available, past frequencies which are relevant. Because probability
quantifies the degree of belief at a given instant, it is not
measurable. Whatever will happen later cannot modify the
probability which was assessed before. It can only influence future
assessments of the probability of other events. Future frequency is
an uncertain number (or ``random variable''), which can assume a
set of values, to each of which we assign a degree of belief. 

\section{Bayesian inference}\label{sec:inference}

Let us consider again the case of an urn containing 70\% white
balls. Imagine that we have made 
$N_0$ extractions out of $N$ total, and have observed the
relative frequency of white balls to be
$f_{W_0}$. It is clear that, given perfect knowledge about the
composition of the urn, all probabilistic considerations about the
remaining $N-N_0$ extractions will be the analogues of those 
initially done for the $N$ extractions.\cite{afterN0} 
The situation changes if we are
uncertain about the composition of the urn. Most likely, after
the first
$N_0$ extractions our beliefs about the result of the remaining
extractions will change. Learning from data is the task of {\it
inference}. This subject is the most interesting part of 
probability theory for physics applications, as we will see in the
following. 

Before attacking the problem formally, it is interesting to
consider what we would intuitively expect. If we have observed only
white balls in the first $N_0$ extractions, we would tend to
believe that the remaining extractions will result in white balls 
much more than the initial 70\%. But it is also clear that this
change of belief would depend on how many extractions have been
made, and how confident we were in our initial 70\%
evaluation. For example, if we had made
only a couple of extractions, or if our prior belief was based on
the information that the urn contains with certainty a percentage
of white balls between 68\% and 72\%, our new belief would not
differ much from the old one. 

\begin{figure}
\begin{center}
\epsfig{file=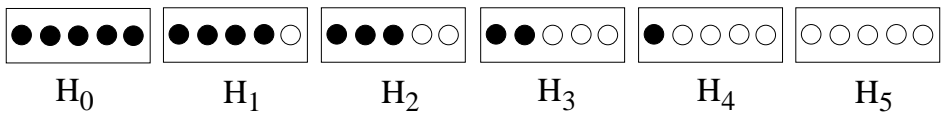,width=0.8\linewidth,clip=}
\end{center}
\caption{\small Six boxes each having a different composition
of black and white balls. One box is chosen at random, then 
its content is inferred by extracting at random a ball from the 
box and reintroducing it inside. What is the probability 
of each box conditioned by all the past observations?
What is the probability of the color of the next ball?}
\label{fig:sei_scatole}
\end{figure}

Now that we have sketched the ingredients which enter an
inferential procedure based on probability calculus, we
illustrate it using an example. Imagine six indistinguishable
boxes with different numbers of black and white balls. The boxes
are labelled
$H_0$, $H_1$, \ldots, $H_5$ according to the number of 
white balls (see Fig.~\ref{fig:sei_scatole}).
Let us choose randomly one of the boxes. 
We are in a status of uncertainty concerning several 
events, the most important of which correspond to the 
following questions.
\begin{enumerate}
\item[(a)]
Which box have we chosen, $H_0$, $H_1$, \ldots, $H_5$?
\item[(b)]
If we extract randomly a ball from the chosen box, will we 
observe a white ($E_W\equiv E_1$) or black ($E_B\equiv E_2$) ball? 
\end{enumerate}
What is certain is that, given the status of information, the
result must be one of the possibilities for each question:
\begin{eqnarray}
\cup_{j=0}^5\, H_j &=& \Omega \\
\cup_{i=1}^2\, E_i &=& \Omega\,.
\end{eqnarray}
In general, we are uncertain about 
all the combinations of $E_i$ and $H_j$: 
$E_W\cap H_0$, $E_W\cap H_1$, \ldots, $E_B\cap H_5$. The 
12 {\it constituents} that we have to consider are not
equiprobable. For example, 
$E_W\cap H_0$ and $E_B\cap H_5$ are impossible.
Because $E_i$ and $H_j$ form complete classes of hypotheses, each
event can be written as a logical sum
of constituents: $E_i=\cup_j(E_i\cap H_j)$,
$H_j=\cup_i(E_i\cap H_j)$. If we remember
that the constituents are by construction
mutually exclusive, we have that 
$P(E_i) = \sum_j P(E_i\cap H_j)$ and a similar sum rule for
$P(H_j)$. If we apply Eq.~(\ref{eq:joint}) to each constituent, we
can express the probability of the events of interest as
\begin{eqnarray}
P(E_i) &=& \sum_j P(E_i\,|\,H_j) P(H_j)
\label{eq:decomp} \\
P(H_j) &=& \sum_i P(H_j\,|\,E_i) P(E_i).
\end{eqnarray}
At this point it is important to model our process of knowledge. 
The $E_i$ play the role of observable {\it effects}: that is,
what we can experience with our senses. The $H_j$ play the role of
{\it physical hypotheses}: they are not directly observable, and in
fact the rule of the game is that we can never look directly inside
a box. In our scheme the 
$H_j$ are the possible {\it causes} of the effects. So the
inference consists in guessing the cause from the effects.\cite{Poincare1}

The experiment consists in extracting balls at random from a
given, but unknown box, and reintroducing it afterward. Our problem
will be that of assessing the probability that the box is a
particular one of
the six boxes shown in Fig.~\ref{fig:sei_scatole}. After we see
the color of the ball, the first intuitive conclusion about the box
content would be that the box that contains more
balls of the same color which has just been extracted is the most
believable. 
This consideration is at the
basis of the {\it maximum likelihood principle}, which is
considered by many people the only (or best) paradigm for making
inferences. However, it is natural to think that the beliefs about
the different causes are constantly updated, and therefore we need a
method for making inferences which goes beyond the maximum
likelihood principle and which takes into account all available
information besides the last experimental observation. 

{}From the previous remark, we can say that 
the aim of a measurement is to update our beliefs about
each cause, given all available information. For example, 
after the first extraction, indicated by $E^{(1)}$, which
could result in either a white ($E_W$) or black ($E_B$) event, 
we will have
$P(H_j\,|\,E^{(1)},I)$; after the first two extractions we have 
 $P(H_j\,|\,E^{(1)},E^{(2)},I)$, and so on. ($I$ 
stands for the all the prior information about the
process and will not be written explicitly in the following.) 
 
Out of the many probabilities we are considering, the easiest ones
to evaluate are the probabilities of observing the different effects
given each cause: $P(E_i\,|\,H_j)$. These probabilities are the
analogue to the response of an apparatus when an
experiment is performed. 
They are technically called {\it likelihoods}, 
because they say how likely the causes produce 
the effects. As for all the probabilities, they can be evaluated in 
several ways. Usually, in real measurements they are evaluated
making use of past frequencies\cite{freq_results}
and some assumptions (beliefs), such as when we state that the
errors are Gaussian distributed. 
In our example they can be evaluated by symmetry arguments, 
and we obtain
\begin{equation}
\left .\begin{array}{l} 
P(E_W\,|\,H_j) = j/5 \\
P(E_B\,|\,H_j) = (5-j)/5 \end{array}\right. .
\label{eq:pEwb}
\end{equation} 

At this point, let us rewrite Eq.~(\ref{eq:joint}) as
\begin{equation}
\frac{P(H_j\,|\,E_i)}{P(H_j)} = \frac{P(E_i\,|\,H_j)}{P(E_i)}\, .
\label{eq:bayes1}
\end{equation}
The meaning of Eq.~(\ref{eq:bayes1}) is that the probability of $H_j$ is
altered by the condition 
$E_i$ in the same ratio by which the probability of $E_i$ is
altered by the condition $H_j$. Therefore, if we know how to
calculate the right-hand-side of Eq.~(\ref{eq:bayes1}), we also
know how to update $P(H_j)$. This ratio is the essence of Bayesian
inference. Clearly $P(E_i)=1/2$ by symmetry, and, hence the
updating ratios are
\begin{equation}
\left . \begin{array}{l} 
\frac{P(H_j\,|\,E_W)}{P(H_j)} = {2\,j/5} \\
\frac{P(H_j\,|\,E_B)}{P(H_j)} = 2\,(5-j)/5 \end{array}\right. .
\label{eq:pHj}
\end{equation} 
If a white ball is observed, all
hypotheses with labels $j \le 2$ become {\it less} credible, while
those with $j\ge 3$ become more credible. The reverse happens
if we observe a black ball. However, the absolute level of
credibility depends also on the initial probability. 

To make this example generally valid,
it is preferable to evaluate $P(E_i)$ in a way
that will be applicable when the symmetry between black and
white is broken, as happens after the observations. We can use 
Eq.~(\ref{eq:decomp}) and obtain, using the equiprobability of the 
box composition: 
\begin{eqnarray}
P(E_i) &=& \sum_{j=0}^5 P(H_j) P(E_i\,|\,H_j) 
 = \frac{1}{6}\times\left(\frac{0+1+2+3+4+5}{5}\right) 
 =\frac{1}{2}\,.
\label{eq:PEi}
\end{eqnarray}
This formula makes explicit our intuitive equal beliefs about black 
and white balls. They depend on the information about
the six boxes.

We can now put all the ingredients together. From 
Eq.~(\ref{eq:bayes1}), using Eqs.~(\ref{eq:decomp}) and 
(\ref{eq:joint}),
we find
\begin{eqnarray}
\label{noness}
P(H_j\,|\,E_i)& = & \frac{P(E_i\,|\,H_j) P(H_j)}
 {\sum_j P(E_i\,|\,H_j) P(H_j)}\,. 
\end{eqnarray} 
The latter formula represents the standard way of writing {\it
Bayes' theorem}. We see that the denominator in Eq.~(\ref{noness})
is just a normalization factor such that
$\sum_j P(H_j\,|\,E_i) =1$. Neglecting the normalization factor
and rewriting $P(H_j)$ as $P_0(H_j)$ to indicate that this
probability is the probability before the observations, we obtain:
\begin{eqnarray}
P(H_j\,|\,E_i) &\propto& P(E_i\,|\,H_j)\, P_0(H_j)\,, \\
\noalign{\noindent or}
\mbox{posterior} &\propto& \mbox{likelihood}\times 
 \mbox{prior}\,.
\label{eq:bayes_scheme}
\end{eqnarray}

Bayes' theorem is simply a compact representation of what has been
done in the previous steps. This point is an important one and is
often misunderstood by those who see Bayesian inference as a kind of
credo or some strange mathematical formalism. Bayes' theorem is a
formal tool for updating beliefs using logic instead of only
intuition. Indeed, we can show that in many simple problems
intuition is qualitatively in agreement with the formal result of
Bayesian inference.\cite{dagocern} But in more complex problems,
intuition might not be enough, and formal guidance becomes crucial.

Table~\ref{tab:sei_scatole} shows the results of a simulated
experiment where the box $H_1$ was extracted (this information was
not available to the analysis program). The second column gives
the result of the first five extractions, together with the
accumulated score in the form $(N_W,N_B)$. After the fifth
extraction, only the score is given. All other columns are
self-explanatory or will be illustrated below. The probabilities
$P(H_j\,|\,I_k)$ are calculated\cite{scores} by iterating Bayes'
theorem: the priors of the present inference are equal to the finals
of the previous one:
\begin{equation}
P(H_j\,|\,I_k) = \frac{P(E^{(k)}\,|\,H_j) \, P(H_j\,|\,I_{k-1})}
 {\sum_l P(E^{(k)}\,|\,H_l) \, P(H_l\,|\,I_{k-1})}, 
\label{eq:iterations}
\end{equation}
where $E^{(k)}$ refers to the $k$th extraction, the $P(E^{(k)}|H_j)$
are given by Eq.~(\ref{eq:pEwb}), and the $P(H_j|I_{k-1})$ are
given by the entries in the previous row of
Table~\ref{tab:sei_scatole}. 

Table~\ref{tab:sei_scatole} shows how the beliefs about the box
composition change with the observations. Note how the hypotheses
which are incompatible with at least one observation are
``falsified'' forever. But, after some observations, all the other
unfalsified hypotheses are not equally likely. This result shows
that probabilistic inference is much more natural and powerful than
Popper's simpler scheme of falsification.\cite{Popper} After
approximately 50 trials, we are practically sure to have obtained
$H_1$, but are never certain. Similarly, we cannot tell that 
$H_2$, $H_3$ and $H_4$ are ruled out. They are simply extremely
unlikely.

Table~\ref{tab:sei_scatole} also shows, as indicated by
$P(E_W\,|\,I_k)$, the belief of obtaining a white ball in the next
extraction (it should be, more precisely,
indicated by $P(E_W(k+1)\,|\,I_k)$). 
They are evaluated applying Eq.~(\ref{eq:PEi}) using $P(H_j) =
P(H_j|I_k)$. After some initial fluctuation, $P(E_W\,|\,I_k)$ converges
to 20\%, consistent with the fact that we assign the highest belief to
$H_1$, which has a 20\% content of white balls. It is interesting
to note that $P(E_W\,|\,I_k)$ is always greater than 20\%. This
result is consistent with the fact that $H_0$ is ruled out at the
first extraction, and hence only boxes with at least 20\% white
balls are considered.

For comparison, Table~\ref{tab:sei_scatole} also gives the observed
relative frequency of white balls, $f(E_W)$. This frequency could
be used as an alternative way of assessing probability. We see that
the convergence to 20\% is much slower than that calculated by
Bayesian inference. Moreover, there are fluctuations below 20\%,
inconsistent with the fact that a white ball percentage below 20\%
 has been proved
impossible. The reason why the Bayesian method works better than
the frequency method is that the latter does not take into account
all of the available information. This problem is a general one
with frequentist methods, which are based on hidden assumptions
of which the user is often unaware. The effect is that practitioners
using frequentist methods often solve problems different than
what they had in mind. For example, in this case the frequency
solution corresponds to a problem with a very large number of
boxes with a white ball percentage ranging almost continuously
from 0 to 100. Clearly a different problem. 

\begin{table}[p]
\begin{center}
\begin{tabular}{|c|c|lccccc|ll|}\hline
 & & &&&&&& & \\
\ trial\ \ & & \multicolumn{6}{c|}{Probability of the hypotheses
$P(H_j|I_k)$} &
$P(E_W\,|\,I_k)$ 
 & $f(E_W)$\ \ \ \\ 
$k$ & $E^{(k)}$ & $H_0$ & $H_1$ & $H_2$ & $H_3$ & $H_4$ & $H_5$ &&\\
 & (score)\mbox{\hspace{6.0mm}} & &&&&&& & \\
 & & &&&&&& & \\
\hline
 & & &&&&&& & \\
0 & -- & 0.167 & 0.167 & 0.167 & 0.167 & 0.167 & 0.167\ \  & 0.50 & --\\
 & & &&&&&& & \\
1 & $E_W$ & 0 & 0.067 & 0.133 & 0.200 & 0.267 & 0.333 & 0.73 & 1\\
& (1,0) & & & & & & & & \\
 & & &&&&&& & \\
2 & $E_B$ & 0 & 0.200 & 0.300 & 0.300 & 0.200 & 0 & 0.50 & 0.50\\
& (1,1) & & & & & & & & \\
 & & &&&&&& & \\
3 & $E_B$ & 0 & 0.320 & 0.360 & 0.240 & 0.080 & 0 & 0.42 & 0.33\\
& (1,2) & & & & & & & & \\
 & & &&&&&& & \\
4 & $E_B$ & 0 & 0.438 & 0.370 & 0.164 & 0.027 & 0 & 0.35 & 0.25\\
& (1,3) & & & & & & & & \\
 & & &&&&&& & \\
5 & $E_W$ & 0 & 0.246 & 0.415 & 0.277 & 0.062 & 0 & 0.43 & 0.40\\
 & (2,3) & & & & & & & & \\
 & & &&&&&& & \\
10 & (3,7) & 0 & 0.438 & 0.468 & 0.092 & 0.002 & 0 & 0.33 & 0.30\\
 & & &&&&&& & \\
20 & (6,14) & 0 & 0.458 & 0.522 & 0.020 & $\approx 10^{-5}$
&0 & 0.31 &0.30\\ 
 & & &&&&&& & \\
30 & (7,23) & 0 & 0.854 & 0.146 & $\approx 10^{-4}$ & 
 $\approx 10^{-10}$ & 0 & 0.229& 0.233 \\
 & & &&&&&& & \\
40 & (9,31) & 0 & 0.936 & 0.064 & $\approx 10^{-5}$ &
 $\approx 10^{-13}$ & 0 
 & 0.213& 0.225 \\ 
 & & &&&&&& & \\
50 & (9,41) & 0 & 0.9962& 0.004 & $\approx 10^{-8}$ & 
 $\approx 10^{-19}$& 0 
 & 0.2008& 0.180 \\ 
 & & &&&&&& & \\
60 & (11,49)& 0 & 0.9985& $0.002$ & 
 $\approx 10^{-10}$ & $\approx 10^{-23}$ & 0 
 &0.2003& 0.183 \\
 & & &&&&&& & \\
70 & (11,59)& 0 & 0.9999& $\approx 10^{-4}$ & 
 $\approx 10^{-13}$ & $\approx 10^{-29}$& 0 
 &0.20002 &0.157 \\
 & & &&&&&& & \\
80 & (12,68)& 0 & 1.0000& $\approx 10^{-5}$ & 
 $\approx 10^{-15}$ & $\approx 10^{-34}$ & 0 
 & 0.200003&0.176 \\
 & & &&&&&& & \\
90 & (15,75)& 0 & 1.0000& $ \approx 10^{-5}$ & 
 $\approx 10^{-16}$ & $\approx 10^{-36}$ & 0 
 & 0.200003&0.188 \\
 & & &&&&&& & \\
100 & (18,82)& 0 & 1.0000& $\approx 10^{-5}$ & 
 $\approx 10^{-16}$ &$\approx 10^{-39}$ & 0 
 & 0.200003 &0.180 \\
 & & &&&&&& & \\
\hline
\end{tabular}
\end{center}
\caption{Results of a simulated experiment in which a box is
selected at random (it happens to be $H_1$) and balls are
extracted and then reintroduced. The analysis program
guesses the box content and the probability of having a white ball
in a future extraction, $P(E_W\,|\,I_k)$. This probability is also
compared to the observed relative frequency of the white balls,
$f(E_W)$.}
\label{tab:sei_scatole}
\end{table}

Coming back to the probability of the different boxes, the 
difference between the Bayesian and frequentist solution 
is not only matter of quantity, but of quality. 
In the latter approach the concept of probability of hypotheses,
a concept very natural to physicists,\cite{Poincare2}
is not defined, and therefore no direct comparison between Bayesian
and frequentist results is possible.
Nevertheless, frequentist methods deal with
hypotheses using the well known procedure of hypothesis tests, in
which a null hypothesis is accepted, or rejected with a certain level
of significance.\cite{htests} 
 Unfortunately, this procedure
is a major source of confusion among practitioners and causes
severely misleading scientific conclusions.\cite{tests}

As a final remark concerning the six box problem, imagine
changing the method of preparation of the boxes. For example, we
could have a large bag containing in equal proportion black and
white balls. We select at random five balls, and without looking at
them we introduce them in the box. Then the game goes on as
before. Clearly the initial beliefs about the box compositions are 
now different, as they can be calculated from the binomial 
distribution: 
\begin{equation}
P_0(H_j) = 
\left(\!\!\begin{array}{c} 5 \\ j
 \end{array}\!\!\right)
\frac{1}{2^5}\,.
\end{equation}
Balanced compositions are more likely than those 
containing balls of the same color. Therefore, even after
the first extraction, the most favored box composition 
will not be that having all balls of the extracted color. This
influence of the conclusions from the prior knowledge is absolutely
reasonable and is mostly important when the number of
extractions is low. It becomes negligible and then disappears
asymptotically when the amount of experimental data is very large. 
Bayesian inference balances in an automatic way the contributions
of experimental evidence and prior knowledge. 

\section{Measurement uncertainty} 

Let us move to the application of Bayesian inference to measurement
uncertainty. Conceptually, it is the same as in the six box
example, except that in most cases 
true values and, as an approximation, 
effects may assume continuous real
values (strictly speaking, effects are by nature discrete).
Let us call $\mu$ the true value and $X$ the observation. Because we are
dealing with continuous quantities, we must use probability density
functions. The function
$f(\mu\,|\,I)$ describes the uncertainty about $\mu$
given the status of information $I$; $f(x,\mu\,|\,I)$ describes the
simultaneous uncertainty about the possible outcome of the
experiment and the true value; $f(x\,|\,\mu,I)$ is related to the
performance of the experiment, as it 
describes the uncertainty about 
the outcome of the experiment under the hypothesis that $\mu$ has 
a particular value; and finally $f(\mu\,|\,x,I)$ is the result of a
measurement, and describes the uncertainty about $\mu$ updated by 
the observation $X=x$.

We could follow the same logical steps sketched for 
the six box example and arrive at an analogous formulation 
for Bayes' theorem, namely
\begin{equation}
f(\mu\,|\,x,I) \propto f(x\,|\,\mu,I) \, f(\mu\,|\,I). 
\end{equation} 
Using the symbol $f_0(\mu)$ for the prior probability density and
assuming
$I$ to be implicit, we have the more compact formula
\begin{equation}
f(\mu\,|\,x) \propto f(x\,|\,\mu) \, f_0(\mu). 
\end{equation} 
Obviously, in this case the normalization denominator is given by
the integral $\! \int\! f(x\,|\,\mu) \, f_0(\mu)\,d\mu$,
integrated over all possible values of $\mu$. 

As an example, consider a detector 
characterized by Gaussian response, that is,
\begin{equation}
f(x\,|\,\mu) = \frac{1}{\sqrt{2\,\pi}\,\sigma}\,
 e^{-(x-\mu)^2/2\,\sigma^2}\,.
\end{equation}
In practice (at least in routine measurements) 
the width of the response 
around the true value $\sigma$ is much narrower than our
uncertainty about $\mu$. For example, if 
the temperature in a room is measured, we would
choose a thermometer which has a $\sigma$ of the order of a degree 
or better; otherwise, we do not obtain a better estimate of the
temperature than what can be inferred from our physiological
feeling. Without going into mathematical proofs, it is plausible 
that if the width of the prior probability density is much larger
than
$\sigma$, the prior probability density acts as a 
constant:\cite{priors}
\begin{equation}
f(\mu\,|\,x) = \frac{ (2\pi\sigma^2)^{-1/2}\,
 e^{-(x-\mu)^2/2\,\sigma^2} \, k}
 {\int_{-\infty}^{+\infty} (2\pi\sigma^2)^{-1/2} \,
 e^{-(x-\mu)^2/2\,\sigma^2} \, k\,d\mu},
\end{equation}
where $k$ is a constant.
Because the integrand is symmetric in $x$ and $\mu$, we obtain: 
\begin{equation}
f(\mu\,|\,x) = \frac{1}{\sqrt{2\,\pi}\,\sigma}\,
 e^{-(\mu-x)^2/2\,\sigma^2} .
\label{eq:res_mu} 
\end{equation}
Note the inverted positions of $\mu$ and $x$ in the exponent, to
remind us that $\mu$ is now the random variable (uncertain number), 
and $x$ a parameter of the distribution. The probability of $\mu$ is
concentrated around the observed value, described by a Gaussian
probability distribution with a standard deviation
$\sigma$. The function $f(\mu\,|\,x)$ contains the complete status
of uncertainty, from which an infinite number of probabilistic
statements about $\mu$ can be calculated. For example, 
if we believe that the detector response is Gaussian
and that $x$ has been observed, then we {\it must} 
attribute a 68\% probability to $\mu$ to be
in the interval $x-\sigma\le\mu\le x+\sigma$, 
 95\% to be within $x-2\,\sigma\le\mu\le
x+2\,\sigma$, and so on.\cite{inversion} 

Although it was not explicitly written in Eq.~(\ref{eq:res_mu}),
we understand that this result depends on all available knowledge
concerning the experiment, including calibration constants,
influence parameters (temperature, pressure, etc), noise, and so
on. In physics jargon, we say, ``it depends on systematic 
effects.'' Let us call all these physical quantities on which the
result can depend {\it influence parameters} and indicate them by
$h_i$. For simplicity, let us assume that each influence parameter 
can assume continuous values. Generally, we are also in a status of
uncertainty about the exact value of these parameters. Because the
uncertainty about one of these quantities could depend on knowledge
about the others, we must consider the general case of a joint
probability density function
$f(\hv) \equiv f(h_1, h_2, \ldots, h_n)$. 
Therefore, the Bayes formula is written, more precisely, as 
\begin{equation}
f(\mu\,|\,x,\hv) \propto 
f(x\,|\, \mu,\hv)\, f_0(\mu)\,. 
\end{equation}
Probability theory tells us how to get rid of the uncertain
influence parameters. We have to make a weighted average over the
possibilities for
$\hv$, with the weight given by how much we believe in
each possibility. Specifically, 
\begin{equation}
f(\mu\,|\,x) = \! \int \! f(\mu\,|\,x,\hv) \, 
f(\hv)\,d\hv\,. 
\label{eq:syst}
\end{equation}
We now have a method of handling uncertainty due to systematic
errors which is very intuitive and does not introduce {\it ad hoc}
ingredients into the theory. There is no well defined and
consistent solution using other approaches.\cite{guide}

As an example, consider a single 
calibration constant related to a scale offset $Z$. 
If the calibration 
had been done, then we believe $Z$ to be around zero, with a
standard uncertainty of $\sigma_z$. 
Let us model this uncertainty by a Gaussian:
\begin{equation}
f_0(z) = \frac{1}{\sqrt{2\,\pi}\,\sigma_z}\,
 e^{-z^2/2\,\sigma_z^2}\,.
\end{equation}
The $z$ dependent likelihood is now 
\begin{equation}
f(x\,|\,\mu,z) = \frac{1}{\sqrt{2\,\pi}\,\sigma}\,
 e^{-(x-(\mu+z))^2/2\,\sigma^2}\,.
\end{equation}
Taking again a constant for the prior probability density for
$\mu$, we have the following inference on $\mu$ conditioned by 
the observed value $x$ and the unknown value $z$:
\begin{equation}
f(\mu\,|\,x,z) = \frac{1}{\sqrt{2\,\pi}\,\sigma}\,
 e^{-(\mu-(x-z))^2/2\,\sigma^2}\,.
\end{equation}
Applying Eq.~(\ref{eq:syst}) we have
\begin{eqnarray}
 f(\mu\,|\,x)
 &=& \! \int_{-\infty}^{+\infty} \!
 \frac{1}{\sqrt{2\,\pi}\,\sigma}\,
 e^{-(\mu-(x-z))^2/2\,\sigma^2}\,
 \frac{1}{\sqrt{2\,\pi}\,\sigma_z}\,
 e^{-z^2/2\,\sigma_z^2}\,dz\,, 
\end{eqnarray}
from which we obtain
\begin{equation}
f(\mu\,|\,x) = \frac{1}{\sqrt{2\,\pi}\,\sqrt{\sigma^2+\sigma_z^2}}\,
 e^{-(\mu-x)^2/2\,(\sigma^2+\sigma_z^2)}\,.
\end{equation}
The probability density function which describes $\mu$ is still
centered around the observed value $x$, but with a standard
deviation which is the quadratic combination of $\sigma$ and
$\sigma_z$. This result is one of the suggested ``prescriptions''
for combining statistical and systematic ``errors''
used by researchers.\cite{errors} In the Bayesian inference it is
just a theorem, with all assumptions clearly stated. Another
interesting property of Bayesian inference is that, when it is
applied to a multidimensional problem, that is, inferring
simultaneously many true quantities from the same set of data with
the same instruments, we obtain a joint distribution $f(\mu_1,
\mu_2,
\ldots, \mu_m\,|\,\mbox{data})$ which also contains the detailed
information about correlations. For further examples, as well as
for approximation methods to be used in everyday applications, see
Ref.~\ref{bib:dagocern}.

As a final remark on measurement uncertainty, let us consider
again the Bayesian inferential framework sketched by
Eq.~(\ref{eq:bayes_scheme}), which is often summarized by the
motto {\it learning by experience}. According to my experience in
teaching, the Bayesian spirit not only shows the correct way of
making inferences, but also gives guidance in the teaching of
laboratory courses. Equation~(\ref{eq:bayes_scheme}) means that
scientific conclusions
depend both on likelihood and prior information. The likelihood describes
the status of knowledge concerning instrumentation, 
environment conditions,
and influence factors, experimenter's contribution, etc. 
Good prior information means a good knowledge of the studied
phenomenology. The importance of these two contributions is well
known to good experimenters. The balance of the two
contributions allows researchers to accept a result, compare it
critically  with others, repeat measurements if needed, calibrate
the instruments, and finally produce useful results for the
scientific community. My recommendation\cite{N1094} is to teach the
theory of measurement uncertainty only after students have
experienced by themselves these aspects of experimentation, and
have learned in parallel the language of probability, the only
language on which a consistent theory of uncertainty can be based.

\section{Summary}
Subjective probability is based on the idea that probability is
related to the status of uncertainty and not (only) to the outcome
of repeated experiments. This point of view, which corresponds to
the original meaning of ``probable,'' was the one to which
Bayes, Bernoulli, Gauss,\cite{Gauss1} Hume, Laplace, and others, 
subscribed.\cite{frequentists} This point of
view is well expressed by the following words of Poincar\'e, ``If
we were not ignorant, there would be no probability, there could
only be certainty. But our ignorance cannot be absolute, for then
there would be no longer any probability at all. Thus the problems
of probability may be classed according to the greater or less depth
of our ignorance.''\cite{Poincare}

The concept of probability is kept separate from the evaluation
rules, and, as a consequence, this approach becomes the most general
one, applicable also to those problems in which it is impossible to
make an inventory of possible and favorable equiprobable cases, or
to repeat the experiment under the same conditions (those problems
are the most interesting ones in real life and research
applications). The other approaches are recovered, as particular
evaluation rules, if the limiting conditions on which they are
based hold.

As far as physics applications are concerned, the importance of the
subjectivist approach stems from the fact that it is the only
approach which allows us to speak in the most general way about
the probability of hypotheses and true values, concepts which
correspond to the natural reasoning of physicists. As a
consequence, it is possible to build a consistent inferential
framework in which the language remains that of probability. This
framework is called Bayesian statistics, because of the crucial
role of Bayes' theorem in updating probabilities in the light of
new experimental facts using the rules of logics. Subjective
ingredients of the inference, unavoidable because researchers do
not share the same status of information, are not hidden with the
hope of obtaining objective inferences, but are optimally
incorporated in the inferential framework. Hence, the prior
dependence of the inference should not be seen as a weak point of
the theory. On the contrary, it obliges practitioners to consider
and state clearly the hypotheses which enter the inference and to
take personal responsibility for the result. In any case, prior
information and evidence provided by the data are properly balanced
by Bayes' theorem, and the result is in qualitative agreement with
what we would expect rationally. Priors dominate if the data is
missing or of poor quality or if the hypothesis favored by the data
alone is difficult to believe. They become uninfluential for routine
high accuracy measurements, or when the evidence provided by the
data in favor of a new hypothesis is so strong that physicists are
obliged to remove deeply rooted ideas. 

The adjectives ``subjective'' and ``Bayesian'' are not really
necessary, and sometimes give the impression that they have some
esoteric meaning. As has been mentioned several times, the intent
is to have a theory of uncertainty in which ``probability'' has the
same meaning for everybody, precisely that meaning which the human
mind has naturally developed. Therefore, I would rather call
these methods {\it probabilistic}. The appellatives ``subjective''
and ``Bayesian'' should be considered temporary, in
contraposition to the conventional methods which are at present
better known.

The status of the art on Bayesian statistics can be found in
Refs.~\ref{bib:BS} and \ref{bib:Kendall}; Ref.~\ref{bib:HU}
provides a general introduction to Bayesian reasoning from an
historical and philosophical perspective.
References~\ref{bib:deFinetti} and \ref{bib:Jeffreys}  are
considered milestones. Many other  references can be found in
Ref.~\ref{bib:dagocern}. Applications in statistical physics can be
found in Refs.~\ref{bib:Jaynes},
\ref{bib:RS}, \ref{bib:MaxEnt}, and \ref{bib:GrNadal}. Finally, as a
starting point for Web navigation, Ref.~\ref{bib:Bayesweb} is
recommended.


\begin{thebibliography}{ref99}
\bibitem{probabile} ``Probable'' comes from Latin and was used
exactly with its contemporary meaning much before a formal
theory of probability was developed.
\bibitem{will} Note how ``will'' does not imply necessarily time
ordering, but a condition of uncertainty concerning something 
that might have been already happened.~\label{bib:will}
\bibitem{torta} The coherence rule is often described in the
following way. Imagine that you assess the value of the
probability, and hence the odds, and then another rational
person chooses the direction of the bet. This situation is similar
to the case where two persons wish to equally divide some goods: one
makes the partition, and the other one has the choice.
\bibitem{conditional} It is obvious that, in an approach in which
probability is always conditional probability, Eq.~(\ref{eq:cond})
cannot ``define'' conditional probability. The interpretation of
Eqs.~(\ref{eq:joint}) and (\ref{eq:cond}) in the subjective
approach is that we are free to assess two of the three
probabilities, but the third one is constrained by coherence. If
the three assignments do not satisfy Eq.~(\ref{eq:cond}), it is
possible to imagine a combination of bets in which one wins or
looses with certainty, depending on the direction of the bets.
Section 8.2 of Ref.~\ref{bib:dagocern} describes an example showing
that the point of view on conditional probability described here is
the same as that intuitively used by researchers.
\bibitem{coins} One could argue that this number can also be
obtained in any other approach, and this argument is formally true. 
The question is how to interpret it. Clearly 23\% is neither a ratio
of the number of favorable cases over the number of equiprobable
cases, nor an evaluation from a long experiment on the relative
frequency of favorable results. Only in the subjective approach is
the result of each step of a probability calculation consistent
with the definition.
\bibitem{Hume2} The concept of probability, well separated by the
evaluation rules, is magnificently expressed in Chapter~6 of Hume's
essay.\cite{Hume}
\bibitem{prevision} The term prevision rather than expected value
is the preferred term of subjectivists. Prevision is a more general
concept than the well known expected value, and can be applied to
uncertain numbers as well as to events. When applied to events,
prevision reduces to probability.
\bibitem{more} The law of large numbers is certainly the most known
and the most misused law of probability. Bernoulli's theorem talks
about probabilities of relative frequencies, and not about a
``limit of relative frequency to probability,'' an expression which
could give the idea of a limit in the usual mathematical sense. The
theorem does not say that if at a certain moment a number in a
lottery has appeared less frequently than what expected from
probability, then it will come out a bit more often in the future
in order to obey the law of large numbers. It does not even justify
the frequency based ``definition'' of probability. As pointed out
by de Finetti,\cite{deFinetti} ``For those who seek to connect the
notion of probability with that of frequency, results which relate
probability and frequency in some way (and especially those results
like the `law of large numbers') play a pivotal role, providing
support for the approach and for the identification of the
concepts. Logically speaking, however, one cannot escape from the
dilemma posed by the fact that the same thing cannot both be
assumed first as a definition and then proved as a theorem; nor can
one avoid the contradiction that arises from a definition which
would assume as certain something that the theorem only states to
be very probable.''
\bibitem{afterN0}
Indicating by the subscript 1 the quantities referring to the 
remaining extractions, we have the obvious result:
$E[f_{W_1}]=p$
and $\sigma(f_{W_1})=\sqrt{p\,(1-p)}/\sqrt{N_1}$. 
Note, however, that the prevision of the relative frequency 
of the entire ensemble is in general different from that 
calculated a priori. Calling $n_1$ the uncertain number of 
favorable results in the next $N_1$ trials, we have
the uncertain frequency 
$f_W = (f_{W_0}\,N_0 + n_1)/N$, and hence 
$E[f_{W}\,|\,N_0]=(f_{W_0}\,N_0+p\,N_1)/N$, 
$\sigma(f_{W}\,|\,N_0)=\sqrt{p\,(1-p)}\,\sqrt{N_1/N}$. It is easy to
understand that, as $N_0$ approaches $N$, we are practically sure
about the overall relative frequency, because it belongs now
to past.
\bibitem{Poincare1} The importance of this reasoning is well
expressed by Poincar\'e: ``\ldots these problems are classified as
{\it probability of causes}, and are the most interesting of all
from their scientific applications.''\cite{Poincare}
\bibitem{scores} This time consuming procedure is not really
needed, although introduced for teaching purposes, and one can use
only the scores. Because the likelihoods in our example do not
depend on the $k$th extraction, if at a certain moment we have
observed $N_W$ white and
$N_B$ black balls (with $N_W+N_B=k$), the iterative application of
Bayes' theorem gives:
\begin{eqnarray*} P(H_j\,|\,I_k) &\propto& P(E_W\,|\,H_j)^{N_W}\,
P(E_B\,|\,H_j)^{N_B}\, P_0(H_j) \\
 & \propto & P(E_W\,|\,H_j)^{N_W}\, [1-P(E_W\,|\,H_j)]^{k-N_W}\,
P_0(H_j) \\ & \propto & P(N_W\,|\,{\cal B}_{P(E_W\,|\,H_j), k})\,
P_0(H_j),
\end{eqnarray*} where $P(N_W\,|\,{\cal B}_{P(E_W\,|\,H_j), k})$
stands for the binomial (${\cal B}$) probability function of
parameters $P(E_W\,|\,H_j)$ and $k$. This result corresponds to the
intuitive idea that, in this problem, 
 the inference should not depend on the order of
the results. The fact that only two numbers ($N_W$ and $N_B$) are 
sufficient to summarize the relevant information for the inference 
is related to the statistical concept of {\it sufficiency}.
Instead, the idea that the $k!/(N_W!\,N_B!)$ possible sequences are considered
a priori equiprobable, though the individual events $E_W^{(i)}$ (not to be
confused with $E_W^{(i)}\,|\,H_j$) are not independent (because the
probability of each event depends on the score of the previous $i-1$ events,
as it is clear from Eqs.~(\ref{eq:PEi}) and (\ref{eq:iterations}) and as can
be easily understood from Table~\ref{tab:sei_scatole}), is related to the
concept of {\it exchangeability}\cite{deFinetti} which we will not consider
here.
\bibitem{Poincare2}Poincar\'e's opinion about the probability of
hypotheses is very enlighting. He calls the problem of assessing
the ``probability of the causes'' (that is, of hypotheses) ``the
essential problem of the experimental method.''\cite{Poincare}
\bibitem{tests} Because this point is rather delicate and touches 
concepts well rooted in all those who are accustomed with standard
statistical methods, it would need a long and careful discussion. I
refer the reader to Ref.~\ref{bib:dagocern} and references therein.
For a short account see also Refs.~\ref{bib:maxent98} and
\ref{bib:BB}. The source of confusion is due to the fact that the
statement, ``the null hypothesis $H_0$ is rejected with a 1\%
significance,'' is interpreted often (from my experience I would
say almost always) as if $H_0$ had only a 1\% chance of being
correct. This mistake is not made only by students, but also by
working scientists.
\bibitem{priors} Obviously, prior knowledge is not always so vague
as to be not influential. If one thinks of two sequential independent
measurements of the same quantity performed with instruments of
(generally speaking) similar quality, the global inference is
obtained by iterating Bayes' theorem, as was seen in the six box
example. The prior of the second inference, i.e. the final of the
first one, has a similar weight of the second data.   
The presence of the priors in the inference is often
considered as a weak point of Bayesian inference. But the criticism
is not justified, because priors play a role which is consistent
with what prior knowledge is expected to do. For an extensive
discussion on this subject see Ref.~\ref{bib:anxiety}.
\bibitem{inversion} This result might seem trivial, because it is
more or less how physicists interpret the results of measurements,
even if they are not aware of Bayesian statistics. This
interpretation is due to the fact that physicists' intuition is
very close to Bayesian reasoning,\cite{maxent98} and probability
inversions of the kind $P(\mu-\sigma \le x \le \mu+\sigma) = 68\%$
implies that $P(x-\sigma \le \mu \le x+\sigma) = 68\%$ are
considered very natural. However, in other approaches this
inversion is arbitrary, although researchers do so intuitively,
with a reasoning described in Refs.~1 and 11. But, unfortunately,
most people are not aware of the implicit assumptions on which this
intuitive probability inversion is based, namely uniform priors and
symmetric likelihood. If these assumptions do not hold, the
numerical results are mistaken.
\bibitem{errors} Errors within quotation marks remind the reader
that error is often used improperly as a synonym for uncertainty.
The metrology organizations, in particular ISO\cite{ISO} and
DIN,\cite{DIN} have done much work to bring some clarification in
the terminology concerning measurement, measurement errors and
measurement uncertainty. The result of this work has been adopted
also by NIST.\cite{NIST}
\end{thebibliography}
\end{document}